# X-ray Emissions in a Multiscale Fluid Model of a Streamer Discharge

Nikolai G. Lehtinen[1] and Nikolai Østgaard[1]

[1]Birkeland Centre for Space Science, University of Bergen, Bergen, Norway

**Abstract** We use a three-specie fluid model of electric discharge in air to simulate streamer evolution from the avalanche-to-streamer transition to the collision of opposite-polarity streamers. We estimate the upper limit on the production of thermal runaway electrons, which is dominant during the second of these processes. More thermal runaways are produced if the ionization due to natural background and photoionization is reduced, due to possibility of creation of higher electric fields at streamer tips. The test-particle simulation shows, however, that these thermal runaway electrons have insufficient energies to become relativistic runaways. The simulations are done in constant uniform background fields of $E_0 = 4$ and 6 MV/m. A simulation was also performed in $E_0 = 2$ MV/m after formation of streamers in 4-MV/m field, in order to approximate the average background field created by ∼1-MV voltage over a ∼1-m electrode gap used in laboratory spark experiments. We conclude that the used fluid model is insufficient to explain X-ray observations during such experiments. We discuss the possible role of mechanisms which were not included in this or previous modeling but may play the deciding role in the electron acceleration and X-ray production during a streamer collision.

## 1. Introduction and Outline

One of the most mysterious phenomena associated with lightning is the intense short energetic emissions called terrestrial gamma ray flashes (TGF; Dwyer et al., 2012; Marisaldi et al., 2013). The spectra of these emissions are compatible with bremsstrahlung from relativistic electrons. The theories of the origin of these electrons employ two mechanisms: (i) multiplication of relativistic electrons in the process of relativistic runaway electron avalanche (RREA) and (ii) acceleration of free electrons from thermal energy, usually referred to as the thermal runaway. The first mechanism requires large voltages (the minimum avalanche length voltage being ∼7.3 MV) and hundreds of meters for sufficient multiplication and the activation of the feedback mechanism (Dwyer, 2003), but the process may proceed in relatively low electric fields, namely, lower than the electric discharge breakdown field or even electric streamer propagation field threshold. Such conditions may exist in thunderclouds but are hard to reproduce in a small-scale laboratory. The second mechanism, on the opposite, requires extremely high electric fields, but the high voltages may be applied over short distances. The conditions are thus feasible to be reproduced in laboratory conditions. It is possible that both mechanisms act at the same time, that is, the RREA may be seeded by the thermal runaway (e.g., Skeltved et al., 2017).

In this paper, we focus on the conditions in a meter-scale megavolt (MV) laboratory discharge. Such experiments show that streamers form complicated branching macroscopic structures of ∼10 cm scale, also containing streamers originating in the space between electrodes. This process is termed *pilots*, and its observations and one-dimensional modeling are presented by Kochkin et al. (2016). The short X-ray emissions occurring during such discharges (Dwyer et al., 2005; Nguyen et al., 2008; Rahman et al., 2008), even though not as energetic as TGF, suggest that such a discharge may serve as a proxy for studying the thermal runaway phenomenon. Moreover, the X-ray emissions come before the development of the leader, which suggests that they are produced by the streamers. In order to tackle the X-ray production problem, we simulate the development of a streamer, starting with an avalanche-to-streamer transition and up to a collision of streamers of opposite polarities.

The avalanche-to-streamer transition was first described by Loeb and Meek (1941). They emphasized the role of photoionization process creating secondary avalanches which provide ionization in the front of a positive streamer, allowing it to propagate. Besides photoionization, we also investigate the role of electron detachment from negative ions. This effect, together with the previously occurring attachment, was neglected by Loeb and Meek (1941). The negative polarity end of the ionized region also eventually becomes a streamer, thus creating a system of two streamers of opposite polarity propagating out in the opposite directions along the external electric field. Note that the avalanche-to-streamer transition that we model occurs in air, without electrodes, which may mimic the streamers originating in the pilots.

If streamers of opposite polarity originate at different points in space and propagate toward each other, this may eventually result in a streamer collision, a process which became a focus of recent streamer research and which we also model in this paper. The timing of observations of X-ray photons in the experiments seems to suggest that colliding streamers may be their source (Kochkin et al., 2012, 2015). Namely, Kochkin et al. (2015, Figure 4) shows that the X-ray pulse occurs during the fourth streamer burst of the prespark phase characterized by enhanced discharging electrode current which lasts ∼200 ns. The streamer collisions with cathode were detected as high-frequency current oscillations (Kochkin et al., 2014), and similar oscillations were within 50 ns of the X-ray detection (Kochkin et al., 2015, Figure 4); however, no single streamer collision has been unambiguously associated with specific X-ray burst or HF (high-frequency) oscillations yet. The average energy of observed X-ray photons was estimated to be ∼86 keV (Carlson et al., 2015). The most feasible and currently accepted hypothesis of production of these photons is by energetic electrons in the process of bremsstrahlung. The source electrons of 200- to 300-keV energies may have also been observed by Østgaard et al. (2016). The source of energetic electrons themselves, however, remains an unanswered topic. It had been proposed that the energetic electrons are created during the streamer collision process due to high electric field between two streamer heads (Cooray et al., 2009). The electrons have to accelerate above ∼100- to 120-eV energy after which the friction force due to collisions starts decreasing from its peak of 26–27 MV/m (Bakhov et al., 2000; Moss et al., 2006) and their subsequent acceleration becomes easier, in a runaway manner.

Previous modeling (Babich & Bochkov, 2017; Ihaddadene & Celestin, 2015; Köhn et al., 2017) suggested that the strong field does not exist long enough to accelerate electrons above runaway threshold, while Luque (2017) suggested that there is an electrostatic wave right after the streamer collision that can accelerate electrons up to 100 keV. We perform similar simulations with an accurate fluid three-specie model which resolves the submicron scale of the thickness of the streamer ionization front. The obtained high electric fields and their lasting times are used to estimate quantitatively the number of thermal runaway electrons produced, by using criteria for acceleration of Bakhov et al. (2000). We will also discuss whether these electrons can become relativistic runaways which are necessary for production of X-rays.

## 2. Model Description
### 2.1. Processes Modeled

The system is described by fluid equations for three charged species, with number densities $n_e$ (electrons), $n_p$ (positive ions), and $n_n$ (negative ions):

$$\frac{\partial n_e}{\partial t} = -\nabla \cdot (\mathbf{v}_d n_e) + \nabla \cdot (D \nabla n_e) + (\nu_i - \nu_a) n_e + \nu_d n_n - \beta_e n_e n_p + s_p + s_{\text{bg}} \quad (1)$$

$$\frac{\partial n_p}{\partial t} = \nu_i n_e - (\beta_e n_e + \beta_n n_n) n_p + s_p + s_{\text{bg}} \quad (2)$$

$$\frac{\partial n_n}{\partial t} = \nu_a n_e - \nu_d n_n - \beta_n n_p n_n \quad (3)$$

$$\nabla \cdot \mathbf{E} = \frac{e}{\varepsilon_0}(n_p - n_e - n_n) \quad (4)$$

where $\mathbf{E} = -\nabla \phi$ is the quasi-static electric field; $\nu_i(E)$, $\nu_a(E)$, and $\nu_d(E)$ are ionization, attachment (both dissociative (two-body) and three-body), and detachment rates, respectively; see Figure 1; $\beta_e$, $\beta_n$ are recombination coefficients with positive ions for electrons and negative ions, respectively; $\mathbf{v}_d = -\mu_e(E)\mathbf{E}$ is the electron drift velocity; and $D(E)$ is the electron diffusion coefficient. The source terms are as follows: $s_p$, the photoionization source (Zheleznyak et al., 1982) whose treatment is described in detail in Appendix A and $s_{\text{bg}} = 10^7$ m$^{-3}$/s is the background ion pair production rate at the ground level due to cosmic rays and mostly radiations from radioactive substances (Goldman & Goldman, 1978, p. 221; Raizer, 1991, p. 141). Most of the coefficients are taken from Morrow and Lowke (1997), while $\nu_d$ is from Luque and Gordillo-Vázquez (2012).

Additional processes, which are not included in this model, and motivation to include them are discussed in section 4.

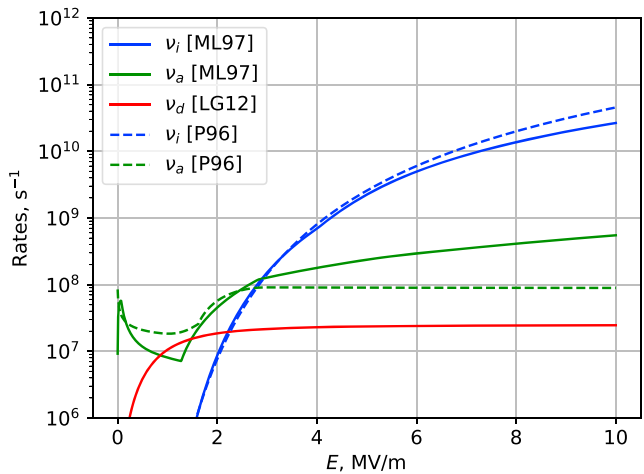

**Figure 1.** Ionization, attachment, and detachment rates (Luque & Gordillo-Vázquez, 2012; Morrow & Lowke, 1997; Pasko, 1996). The attachment includes both dissociative and three-body components.

## 2.2. Numerical Approach

As discussed below (section 4.1), to accurately model the formation, propagation, and collision of streamers, it is necessary to have sufficient spatial resolution, which may be as low as 1 μm. Unfortunately, many modeling efforts use much bigger scale, for example, Ihaddadene and Celestin (2015) use a uniform cylindrical grid with resolution of 8 μm. This may have led to excessive numerical errors, such as numerical diffusion, which could have significantly affected the streamer propagation. The typical streamer radius is 0.1–1 mm (Raizer, 1991) and the length a few centimeters. In order to accurately model it with a uniform grid, an unfeasibly large number of grid points would be required. The solution is to use an adaptive grid which resolves the head of the streamer with smaller grid steps and uses larger grid steps for other parts of streamers. This approach was used by Morrow and Lowke (1997) who modeled a cylindrically symmetric streamer by using a nonuniform grid in $r$ direction expanding exponentially from 10 μm on the axis with a total of 50 grid points over 2 cm and an adaptive grid in $z$ direction with a step 1 μm at the head of the streamer and a larger step size at other locations, with 400 points distributed over 5 cm. Another very promising approach is to use a more sophisticated adaptive grid algorithm as was done in Afivo code (Teunissen & Ebert, 2017), which splits each grid cell halfways along each dimension when higher resolution is required.

We also use an adaptive mesh refinement (AMR) approach but in a slightly different way than those described above. Namely, on a coarse mesh on the main domain we create extensive submeshes over the subdomains that require higher resolutions. This process is repeated with the submeshes. The criterion for refinement is given quantitatively by a number $B_{ij}^{r,z}$ (which may be called *resolution insufficiency* or simply *badness*) which is $> 1$ if the resolution along $r$ or $z$ at point $(i, j)$ is insufficient even to get a qualitatively correct solution, that is, the calculated values have big jumps comparable to the amplitudes of values themselves. For qualitatively correct calculations, we must have $B < B_{min}$ everywhere with $B_{min} \lesssim 1$. Ideally, it should be $B_{min} \ll 1$, but in practice, we choose $B_{min} = 1$–2 or 0.5–1 (the range is needed so that subdomains are not modified at each time step). There are many different ways to calculate $B$ (Li, 2010), which we may try in the future. Currently, we are using the criterion based on the second spatial derivatives of various reaction rates in equations (1)–(3). Namely, for each function $F(\mathbf{r}) = n_e, v_i n_e, v_i n_e n_e n_p$ we determine $\Delta x^0(\mathbf{r})$, the largest acceptable grid step at a given location $\mathbf{r}$ and in given direction $x$. It is calculated as $\Delta x^0 = \sqrt{|F/F''|}$ and is needed only at points where functions $F$ have significant values, for example, $|F| > 0.01 \max |F|$. The badness may then be defined as $B(\mathbf{r}) \equiv \max(\Delta x/\Delta x^0)$, where $\Delta x$ is the actual grid step, and the maximum is taken over different functions $F$ and different directions $x$.

As the streamer propagates, the badness is recalculated and the subdomains need to be modified. We have implemented an algorithm which tries to minimize the number of subdomain modification operations. In particular, subdomains include a margin of otherwise *good* points at their edges, so that the *bad* points do not lie close to the coarser resolution. The implemented subdomain operations include their moving, resizing, splitting, merging, and erasing. When the domains are modified, it is taken care of that the high-resolution content is not lost.

Simulations are done using a 2-D third-order upwind flux-corrected scheme for advection, modified from the 10-point stencil UTOPIA scheme of Leonard and Niknafs (1990) by expanding the stencil to 12 points in order to make it more stable. We also use the flux correction scheme by the same author (Leonard, 1991), which we generalized to two dimensions. The finite difference algorithms are implemented with the help of a custom Python finite-difference package which is available together with the code (see Acknowledgments) and which also includes extensive testing of the advection schemes.

The time stepping is done either with a first-order explicit Euler scheme or with a second-order midpoint scheme, in which reaction rates in equations (1)–(3) for finding the values of $n_{e,p,n}$ at the next time moment $t + \Delta t$ are calculated at the midpoint $t + \Delta t/2$. In practice, decreasing $B_{min}$ (i.e., forcing a better spatial resolution) gave better improvements in accuracy than going from the first-order to the second-order time stepping scheme. The time step $\Delta t$ is limited by the usual stability conditions for discretized diffusion and advection equations, by various reaction rates, and by the Maxwellian charge relaxation rate.

## 3. Results

We model the discharge in a cylindrically symmetric domain of length $L = 15$ mm and radius $R = 2$ mm with uniform external field $E_0$ applied along the $z$ axis. The main domain uses a coarse uniform grid with $\Delta r = \Delta z = 20$ μm. This grid step is clearly insufficient to resolve the streamer propagation. Thus, we proceed to create the refined subdomains, increasing resolution by a factor of 5 at each stage. Namely, the subdomains have resolution of 4 μm and the subsubdomains 0.8 μm. We did not proceed further in the refinement than the second stage. An example of subdivision of the main domain into subdomains and subsubdomains is demonstrated in Figure 2. Note that with the AMR approach, we still have in total fewer grid points than Ihaddadene and Celestin (2015) who modeled a smaller domain of $L = 8$ mm, $R = 1.5$ mm.

The boundary conditions in the main domain are periodic in $z$, so that the same setup can be used both for modeling the avalanche-to-streamer transition and the streamer collisions after the generated positive and negative streamers propagating in opposite $z$ directions wrap around and move to meet each other. We use Neumann boundary condition for $\phi$ on the side surface (i.e., $E_r = 0$ at $r = R$), which allows the field to become high compared to background $E_0$ even when close to the boundary. We note that Dirichlet boundary conditions (i.e., $E_z = 0$ at $r = R$), which were used, for example, by Ihaddadene and Celestin (2015), keep the $E_z$ field uniform close to the boundary, which may be unfavorable to creation of high electric fields during the streamer propagation process. In the future, we may also try the free boundary conditions which effectively place the boundary at infinity. However, we do not expect the results to change significantly, since the radius of the simulated streamer in Figures 3 and 6 never exceeds only ∼25% of the radius of the simulation domain (note that these plots are in log scale). The detailed comparison of various boundary conditions for a spherical charge distribution (Malagón-Romero & Luque, 2018, Figure 2) shows that the fields (inside the sphere) calculated with Neumann boundary conditions are indistinguishable from the exact solution if the radius of the sphere is ≲30% of the radius of the simulation domain. This conclusion may, however, change for an elongated charge system, such as a streamer.

Because of periodicity in $z$, special care must be taken in the code when the refined subdomains wrap around. In the subdomains, the boundary conditions are given by interpolated values of variables in the bigger domain. For solutions of coarser equations, the values from smaller domains are averaged to be used in larger domains.

Parenthetically, we note that, for the chosen type of boundary conditions in the main domain, the solution only exists when the total charge inside this domain is zero, and the electrostatic potential is defined only up to a constant. Thus, besides the conditions at the boundary, one more (gauge) condition has to be enforced.

### 3.1. Avalanche-to-Streamer Transition

We fix the background field of $E_0 = 4$ MV/m and start with ionizing a small volume at the origin ($r, z = 0$). The initial ionization is in a Gaussian profile spherically symmetric region with radius 30 μm and the maximum value of $n_e = n_p = 10^{15}$ m$^{-3}$. Figure 3 shows the time evolution of electron density. The first frame shows the developed avalanche moving downward (direction opposite to the applied $E_0$). When the space charge becomes large enough, the free electrons create positive and negative streamers which propagate in opposite directions. The transition is captured in the second frame of the figure (20 ns), when the positive streamer starts to form. For the process of the avalanche-to-streamer transition to occur, we must start with the field which is higher than the theoretical lower limit $E_b$ for air breakdown (Raizer, 1991 p. 338) defined by $v_i(E_b) = v_a(E_b)$, which gives $E_b \approx 2.8$ MV/m for values used (Morrow & Lowke, 1997). The simplified Meek criterion for the avalanche-to-streamer transition is $d(v_i - v_a)/v_d \approx 20$, where $d$ is the length of the avalanche and $v_d$ is the drift velocity (Raizer, 1991, p. 336), from where we get $d \approx 4.7$ mm for this value of $E_0$. This is consistent with our result shown in Figure 3. The value of $d$ goes to infinity as we get closer to $E_b$, for example, for $E = 3.5$ MV/m we have $d \approx 10.4$ mm, which is already more than half of the length of our simulation domain, so we did not try any lower electric fields.

The avalanche-to-streamer transition process occurs in a different manner, depending on whether we include the detachment process or not. The originally proposed mechanism for this transition (Loeb & Meek, 1941) only names the photoionization as the source of electrons which feed the positive streamer. However, we found that electrons produced by detachment from negative ions which stayed in the trail of avalanching electrons can also provide such a source and the streamer is formed faster than in the case when detachment process is not included in the model (see the ≈3.5-ns delay between the first peaks in Figure 4). The positive streamer is thus formed by concentrating the positive charge into a small volume. Consequently, the positive streamer head diameter, at least initially, is smaller than the negative streamer head diameter. The smaller diameter leads to higher electron concentration in the positive than negative streamer head, an effect which is visible in Figure 3.

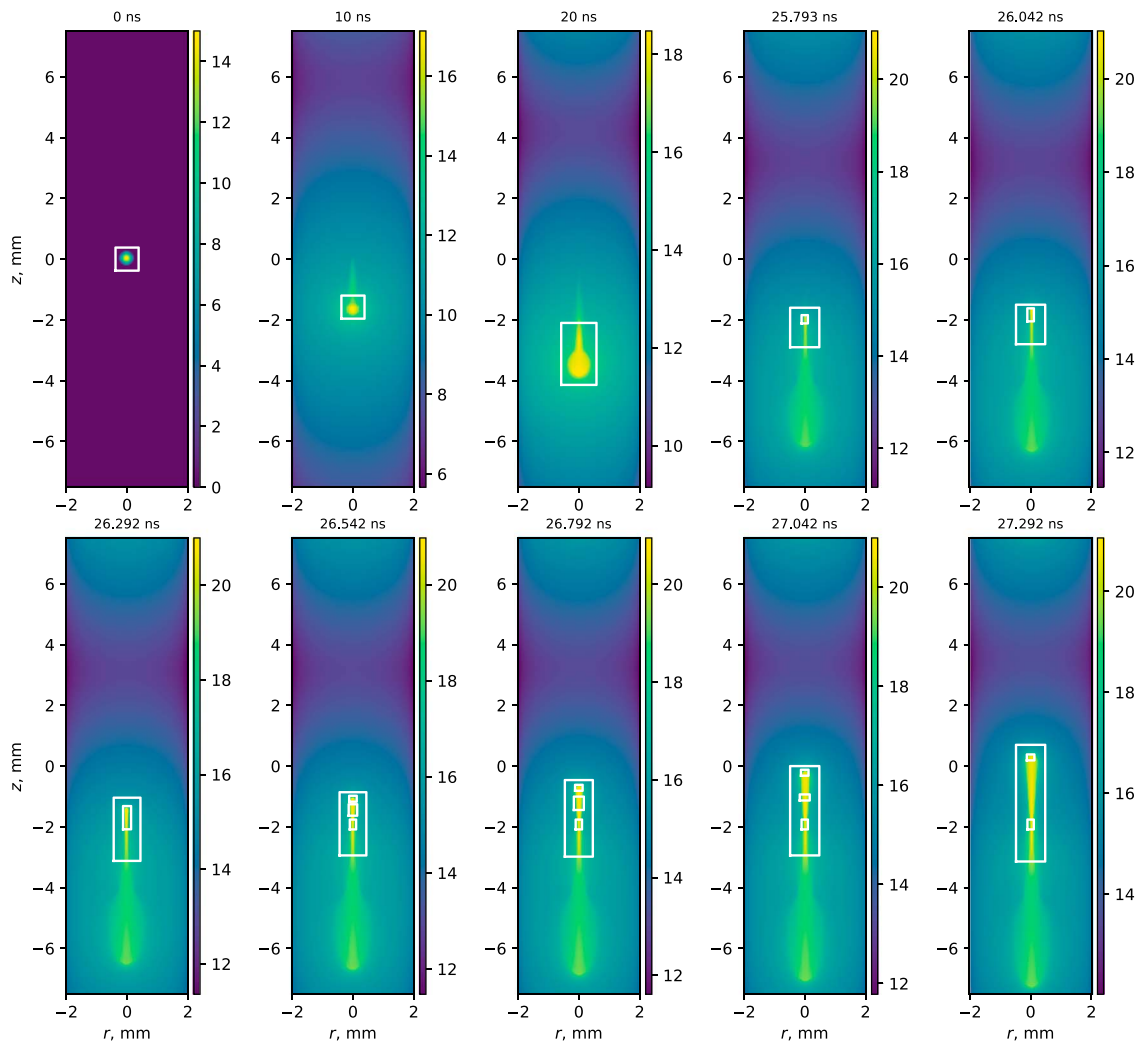

**Figure 2.** An example of adaptive mesh refinement subdivision into a subdomain tree. The 2-D plot is $\log_{10} n_e$ at various moments of simulation, with the refined subdomains and subsubdomains plotted as white rectangles.

### 3.2. Production of Thermal Runaway Electrons

Electrons in air have a peak in dynamic friction function around energy of ~100–120 eV at value ~26–27 MV/m (Bakhov et al., 2000; Moss et al., 2006). When the electric field gets close to this limit, we expect that some of the electrons will become *runaways*, that is, overcome this peak and continue to even higher energies if the electric field in which they propagate allows it.

We consider the process of electron acceleration from thermal energies above this peak modeled by Bakhov et al. (2000). In particular, we use their calculated rates of electrons to reach energy of >4 keV in nitrogen. These electrons will be called *thermal runaway* electrons in this paper, even if they will slow down eventually when they leave the high-field region. Although the calculations were done for nitrogen, the results should also apply to air because air consists of 79% nitrogen, with the rest being mostly molecular oxygen. Oxygen has atomic mass close to that of nitrogen, and therefore, its characteristics for electron propagation are similar. There are some differences, such as in electron attachment, but these are only relevant to electrons with energies well below 100 eV.

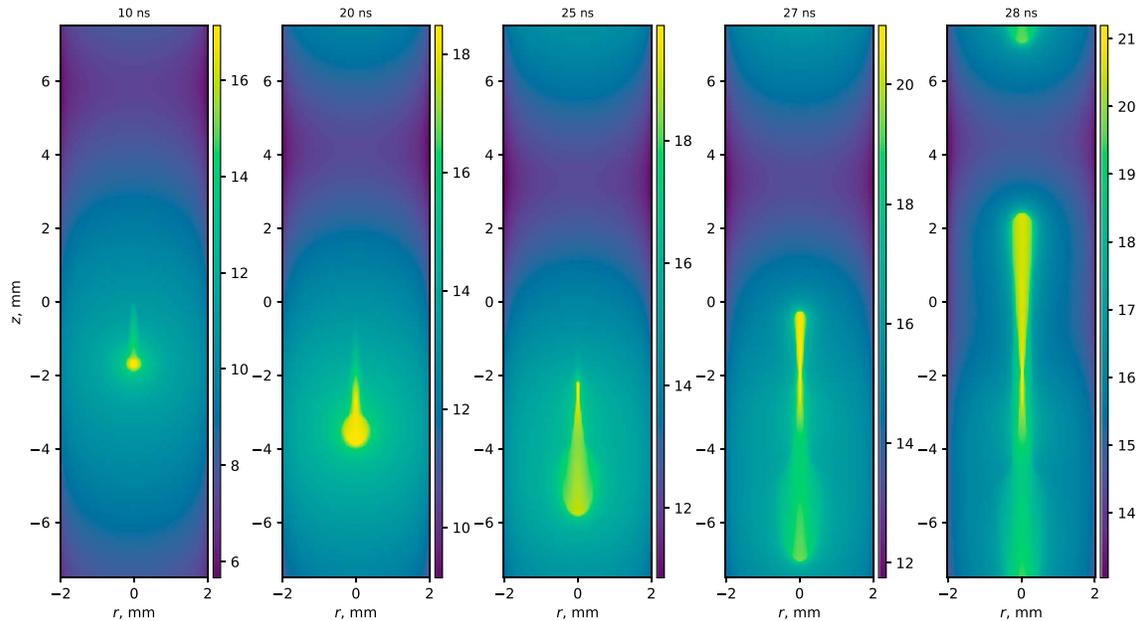

**Figure 3.** Evolution of $\log_{10} n_e$ in the avalanche-to-streamer transition, with the background upward field of $E_0 = 4$ MV/m. We start with small ionization at the origin, which then develops an avalanche in the direction opposite to **E** (downward in this plot). When the electron density becomes sufficiently large, a positive streamer develops upward, while the negative streamer continues to propagate downward. With periodic boundary conditions in the vertical direction, an opposite polarity streamer collision will occur.

The process of conversion of thermal electrons into thermal runaway electrons with energies >4 keV is described by (Bakhov et al., 2000, equation (5))

$$\frac{dn_r}{dt} = \nu_r n_e$$

where $n_r$ is the thermal runaway electron density and $\nu_r$ is the rate of conversion which is dependent on electric field and plotted in Figure 5 of Bakhov et al. (2000). We use an analytical fit for this rate

$$\nu_r(E) = 10^{11-6\exp[-(E/10^6 - 25)/12]} \mathrm{s}^{-1}, \qquad E \text{ in V/m} \qquad (5)$$

This value depends very sensitively on the parameters used in the Monte Carlo simulation of Bakhov et al. (2000). In particular, they demonstrated that $\nu_r$ varies by a factor of ~4 up and down if the used ionization cross section is varied just by ±10%. This justifies the use of the analytical fit (5) in our calculations even though it is somewhat approximate.

The production of thermal runaway electrons in a highly inhomogeneous electric field will be smaller than the production represented by equation (5) because it was calculated for uniform field. This is due to the fact that the electrons may leave the region of high $E$ before accelerating to high enough energies (Kunhardt & Byszewski, 1980). In this work, this effect is included indirectly by limiting the total available energy to the electron, as described in section 3.5.

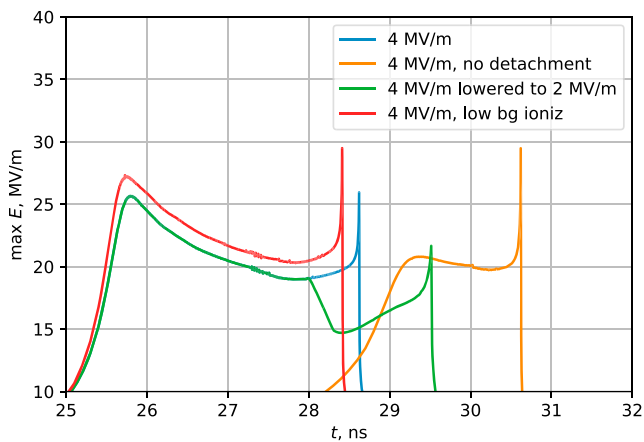

**Figure 4.** The maximum available electric field $E_{\max}$ during the simulation. We also have modeled the case without detachment, the case of the lower background field of 2 MV/m, and the case with lower background ionization (see section 3.4). The field location is at the head of the positive streamer. The first (smaller) peaks correspond to streamer formation and the second (larger) peaks to streamer collision.

The rate $\nu_r$ is small for the values of the fields obtained in the model compared to the time scales during which these fields exist. On the typical nanosecond scale of the processes modeled in this paper, equation (5) suggests that the field of $E \sim 35$–$40$ MV/m is required for significant conversion of thermal electrons into runaways. This led the previous researchers (Babich & Bochkov, 2017; Ihaddadene & Celestin, 2015; Köhn et al., 2017) to conclude that the processes modeled during the streamer collisions are insufficient to explain X-ray emissions observed in experiments (Kochkin et al., 2012, 2015). We will quantify the cumulative number of the thermal runaway electrons as an upper limit on the number of runaway electron produced by integrating the production rate over time and volume:

$$N_r(t) = \int_0^t S_r(t')\,dt', \qquad S_r(t) = \int \nu_r(E) n_e\,dV$$

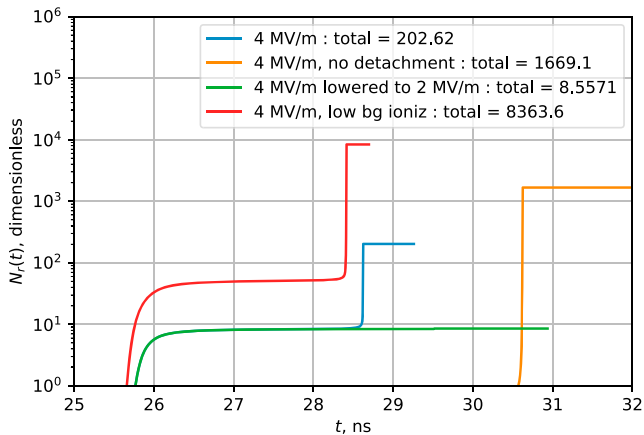

**Figure 5.** The cumulative number of thermal runaway electrons. The final number $N_{r,max}$ is given in the legend. The cases modeled are the same as in Figure 4.

We simulated several cases corresponding to variations in physical parameters, all for background field of $E_0 = 4$ MV/m. For all of them, the maximum electric field $E_{max}(t)$ as a function of time is presented in Figure 4, and the cumulative number of thermal runaways $N_r(t)$ is presented in Figure 5. The highest field in all of these cases is at the head of the positive streamer, which is due to its smaller diameter than that of the negative streamer.

1. *Normal* case when equations (1)–(4) are solved as described in section 2.1.
2. Hypothetical situation when there is no detachment. As discussed above in the context of the avalanche-to-streamer transition, this precludes formation of the positive streamer on the ionization trail left by the initial avalanche. A different mechanism for streamer formation also explains the later start of the high electric fields in Figure 4.
3. The electric field is gradually lowered to 2 MV/m after the streamers are already formed. This is done in order to simulate the background field during streamer collision which is closer to the average background field in a meter-scale laboratory spark experiment. In our simulation, the field decreases linearly starting at $t = 28$ ns over the duration of 0.3 ns. Although the streamers continue propagating, the collision does not create a very high electric field. The streamer propagation is possible because the new background field is still higher than the well-known experimental values of minimum fields required for streamer propagation of $E_{+s} = 0.45$ MV/m for positive streamers and $E_{-s} = 0.75$–$1.25$ MV/m for negative streamers
4. Lower background ionization, with conditions described in section 3.4, to test how the background ionization affects the electric fields. Namely, we considered lower photoionization ($p_q = 30$ mmHg instead of 60 mmHg in equation (A2) and background ionization ($s_{bg} = 10^6$ m$^{-3}$/s instead of $10^7$ m$^{-3}$/s). The lower background ionization, according to Tinsley and Zhou (2006), may correspond to locations over oceans.

Many of these cases show common features. For example, the two peaks in $E_{max}$ correspond to (i) the high field during the initial avalanche-to-streamer transition (taking place over time of $\sim 1$ ns) and (ii) the high field during the very short ($\lesssim 10$ ps) streamer collision process. From Figure 5 we see that most ($>95\%$) of the thermal runaway electron production takes place during the streamer collision.

### 3.3. Case $E_0 = 6$ MV/m

The time evolution of the avalanche-to-streamer transition occurs faster in this case of higher background electric field. The time evolution of electron density is shown in Figure 6, and the maximum field as a function of time is shown in Figure 7. We see that it did not exceed 15.3 MV/m and was even lower during the streamer collision process. We attribute it to the fact that the space between the opposite-polarity streamers was preionized due to high photoionization. The high electron density due to photoionization is visible in Figure 6 as the *fuzziness* surrounding the streamers. The higher ionization outside streamers gave rise to lower field, by the mechanism discussed in section 3.4. Thus, Cooray et al. (2009) mechanism definitely is not able to work in this case.

### 3.4. Role of Background Ionization in the Streamer Fields

The higher electric fields and subsequent higher thermal runaway production for the case without detachment may be explained by the fact that detachment creates additional ionization in front of the streamer, thus increasing conductivity in front of the streamer and leading to lower electric field at the head of the streamer:

$$J = \sigma E \approx \text{const} \Longrightarrow E \downarrow \text{ as } \sigma \uparrow$$

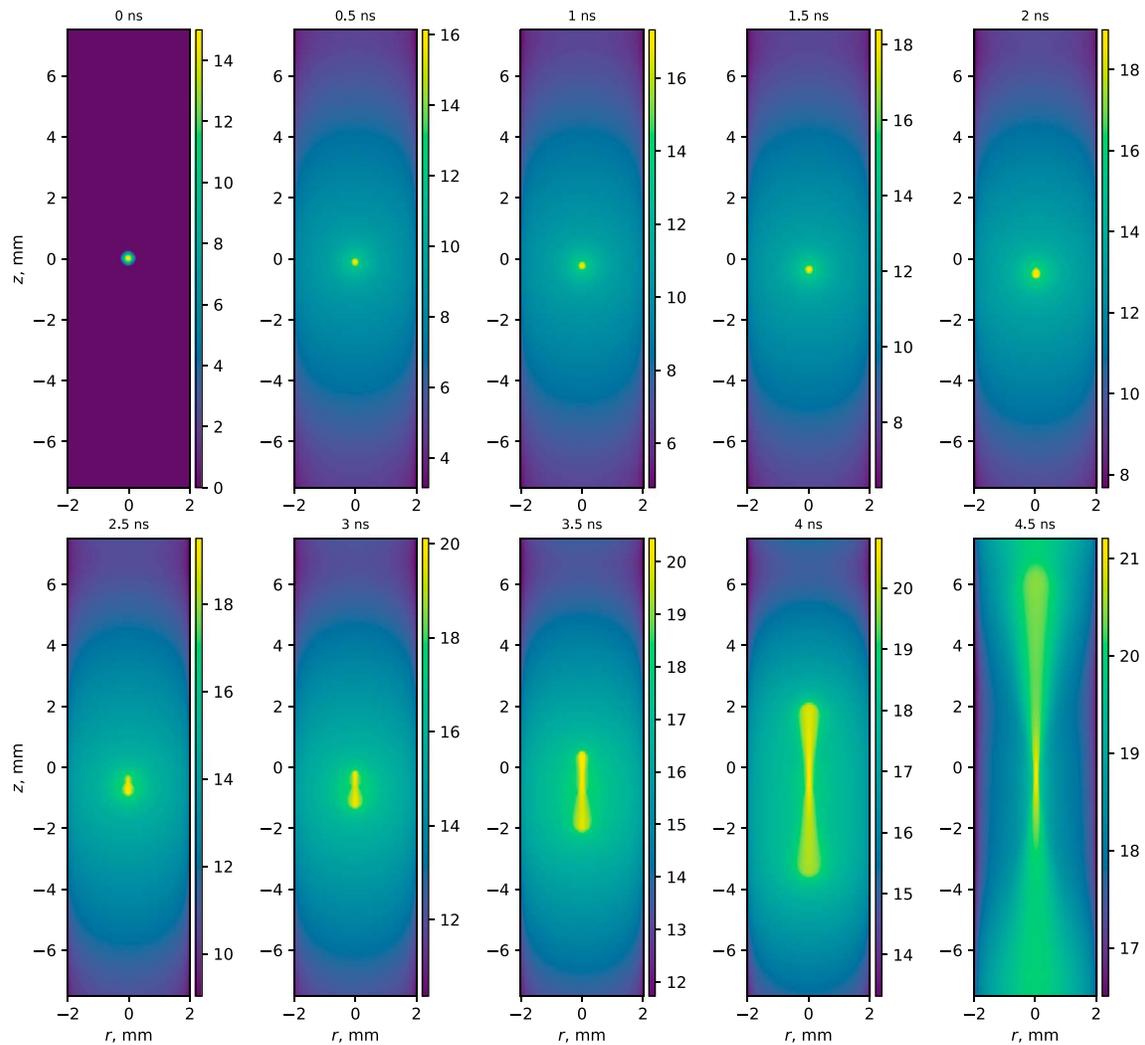

**Figure 6.** Time evolution of $\log_{10} n_e$ for $E_0 = 6$ MV/m.

To check this hypothesis, we performed an additional simulation in which we lowered the photoionization (namely, took $p_q = 30$ mmHg instead of 60 mmHg) and the background ionization ($s_{bg} = 10^6$ m$^{-3}$/s instead of $10^7$ m$^{-3}$/s). The curves for this case are also included in Figures 4 and 5, and indeed, we see that the electric fields are slightly higher in this case (fourth curve) than in the normal case (first curve). This type of background variation may be possible in experimental conditions. If the background ionization is lowered further, then the maximum $E$ and $N_r$ become even higher. In the extreme (unrealistic) case of $p_q = 1$ mmHg and $s_{bg} = 0$, we obtained maximum $E \sim 55$ MV/m and $N_r \sim 10^7$. However, at such high values of $E$ the used expressions for various reaction rates probably become invalid.

The described effect also explains relatively low electric fields in the heads of colliding streamers in the $E_0 = 6$ MV/m case, as the photoionization production there is very high.

### 3.5. Test Particle Acceleration/deceleration

The average field between electrodes in the laboratory spark experiments is of the order of 1 MV/m, which is well over the relativistic runaway threshold of $\approx 0.28$ MV/m (see Appendix C1). However, even if the thermal runaway electrons are created in the region of high $E$, they do not necessarily become relativistic runaways. Such a situation arises if the thermal runaway electron energy is too low so that the friction (see Figure C1) is too high. Then, after an electron leaves the high field region between the streamers (which has small spatial dimensions, ~5 μm), it encounters a region with low $E$, which is insufficient for runaway process and slows down. In order to test this, we perform a test-particle simulation.

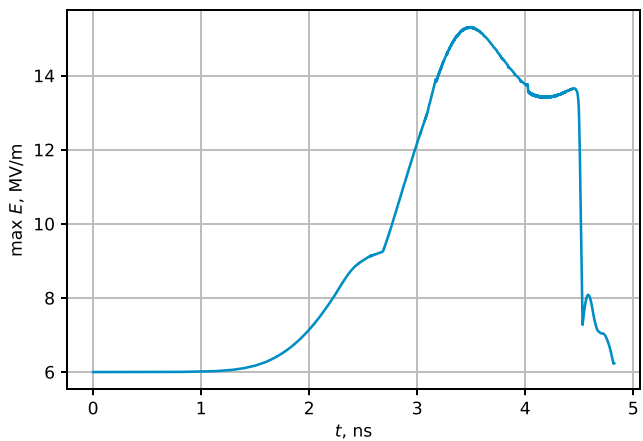

**Figure 7.** Maximum value of $E$ for $E_0 = 6$ MV/m.

To avoid a full Monte Carlo-type simulation, in the present work we represent stochastic motion of test particles by deterministic equations, as outlined in Appendix C. Thus, we solve equations of motion (C2)–(C4) using the calculated electric field. The initial particle energy is chosen to be the maximum available energy, which is defined by the maximum potential drop in regions where the fields exceed 10 MV/m. Since the thermal runaway process does not start before the field reaches about 24 MV/m (Bakhov et al., 2000), this would give us an overestimated initial energy of an electron. However, we will momentarily show that even the overestimated energy is insufficient for these electrons to become relativistic runaways.

We solve equations of motion for the test electrons in one dimension only (along the electric force, i.e., opposite to electric field). A sample electron trajectory is presented in Figure 8, for an electron of maximum available energy generated during the streamer collision, when the thermal runaway electron production rate is the highest. Shown is the normal case for background field of $E_0 = 4$ MV/m. We see that the electron stops after traveling about 1 mm along the electric force. We also show a trajectory of an electron with a minimum energy required to run away. For this particular case, the maximum available initial thermal runaway electron energy is $\mathcal{E}_0 = 9.1$ keV, while the energy necessary for relativistic runaway is $\mathcal{E}_r = 17.1$ keV. There is a finite but very small probability for an electron to become a runaway even if its energy is far below the runaway region boundary. Such probabilities were calculated, for example, by Lehtinen et al. (1999, Figure 6). In the case considered here, the energy is almost twice as small as required for runaway. The probability for such a low-energy electron to become a runaway in a uniform system was estimated using a Monte Carlo model (Lehtinen et al., 1999) to be of the order or less than ~$10^{-3}$, which is less than $1/N_{r,max}$ obtained in our calculation. To determine the probability in a nonuniform field, Monte Carlo calculations are required.

We performed similar calculations also for other cases considered in section 3.2. In a system without detachment, the maximum available energy was $\mathcal{E}_0 = 10.3$ keV and the energy necessary for relativistic runaway was $\mathcal{E}_r = 15.5$ keV. In the case with low background ionization ($p_q = 30$ mmHg instead of 60 mmHg; $s_{bg} = 10^6$ m$^{-3}$/s instead of $10^7$ m$^{-3}$/s), we got $\mathcal{E}_0 = 9.9$ keV and $\mathcal{E}_r = 15.9$ keV. Although in these cases the ratio $\mathcal{E}_0/\mathcal{E}_r$ is slightly closer to 1, we did not identify the trend of its increase when the background ionization is reduced even further. Namely, for the extreme case considered in section 3.4 ($p_q = 1$ mmHg and $s_{bg} = 0$), we obtained $\mathcal{E}_0 = 6.8$ keV and $\mathcal{E}_r = 15.1$ keV.

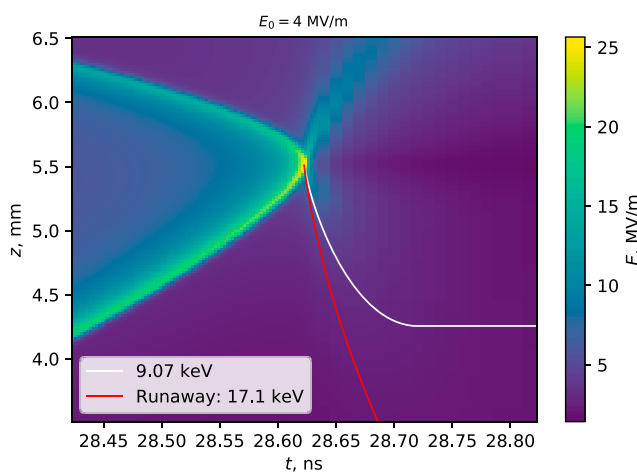

**Figure 8.** Trajectories of two electrons generated during a streamer collision ($E_0 = 4$ MV/m case): with maximum available energy due to potential drop (9.07 keV) and the minimum energy required for runaway (17.1 keV).

We observe that after the streamer collision there are two electrostatic waves traveling in opposite direction, as if the streamers have gone through each other (the wave traveling along electric field is more visible in the Figure). These waves have occurred also in electromagnetic modeling of streamers by Luque, (2017, Figure 4), who concluded that these waves can accelerate electrons up to 100 keV, starting with 1 keV. However, in our simulations these waves do not overcome the effective friction in order to accelerate the generated electrons. If we also neglect elastic collisions and use equations (C5) and (C6) to describe electron motion, then we also are able to produce relativistic runaway electrons. Monte Carlo calculations of runaway electron avalanche by Lehtinen et al. (1999), however, suggest that the elastic collisions cannot be neglected.

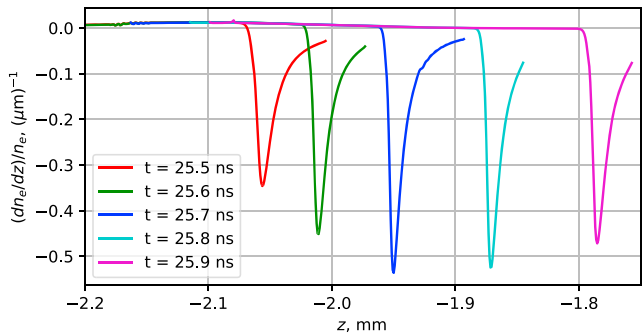

**Figure 9.** Steepness of the electron density gradient $\nabla n_e/n_e$, taken on the axis of the simulation domain for the case of $E_0 = 4$ MV/m around the moment of formation of the positive streamer.

## 4. Discussion and Conclusions

### 4.1. Necessity of Multiscale Approach

In a streamer, there is a length scale (for gradients) present at which the electron speed due to diffusion $\mathbf{v}_D = -D(E)(\nabla n_e)/n_e$ is approximately equal in magnitude to the drift speed $\mathbf{v}_d = -\mu(E)\mathbf{E}$:

$$v_D \approx \frac{D}{d_D} = \mu(E)E$$

$$d_D = \frac{D}{\mu E} = \frac{T_e}{eE} \approx \frac{2}{3}\frac{\mathcal{E}}{eE}$$

where we used the Einstein's relation between the diffusion $D$ and mobility $\mu$; $\mathcal{E}$ is the typical electron energy.

This scale also comes up in the numerical scheme developed by Kulikovsky (1995), who derives a condition on the grid step, namely, $\Delta x_k \ll d_D$. If this condition is not satisfied in that scheme, it essentially becomes a first-order upwind scheme and suffers from excessive numerical diffusion. We point out that the third-order scheme used in this work does not have this property.

We must mention that Naidis (2009) obtained the ratio of diffusion and drift terms in the density change equation (1) to be < 0.1. This was based on simulation results of $T_e = 10$ eV, $E = 10$ MV/m, and the gradient scale $n_e/|\nabla n_e| \equiv l_D > 10$ μm. However, in our calculations we get much steeper gradients, especially in the beginning of the positive streamer propagation (i.e., around $t = 25$ ns in Figure 3). For such sharp gradients, the diffusion term is important. In our model, the AMR adjusts the grid size to the existing gradients, and 0.8 μm resolution was sometimes required to fully resolve the ionization front of a moving streamer, as seen in Figure 2. The steepness of electron density (equal to $1/l_D$) around the time of the positive streamer formation is plotted in Figure 9, which shows that it is possible that $l_D < 2$ μm.

### 4.2. Important Mechanisms Not Included in the Fluid Model

The streamer simulations, such as Liu and Pasko (2004) and Ihaddadene and Celestin (2015), usually do not take into account the nonlocal effects, such as the reactions coefficients (e.g., $v_{i,a}$) being determined not by the local electric field but by the field where the electron distribution was created. The typical length scale of such nonlocality is the energy relaxation scale (Raizer, 1991, p. 19) $\Lambda_u = v_d/v_{\mathcal{E}}$, where $v_{\mathcal{E}}$ is the electron energy loss rate due to collisions with neutrals. These effects were considered by Naidis (1997), in approximation derived by Aleksandrov and Kochetov (1996). The results of (Naidis, 1997, Figure 1a) suggests that electric field at the streamer head is higher by about 30% when these effects are included. This can significantly change the conclusions about X-ray production, because of the highly nonlinear dependence of the thermal runaway rate (5) on the electric field. Moreover, the approximation used by Aleksandrov and Kochetov (1996), namely, $\Lambda_u \ll L$, where $L$ is the typical scale of nonuniformity, may not be completely valid if the streamer ionization front thickness has sub-μm scale, as values of $\Lambda_u$ shown in Figure B1 suggest. The current efforts to include nonlocal effects using particle-in-cell (PIC) modeling by Köhn et al. (2017), however, give similar results for the X-ray production to the fluid model approach.

Another effect which is not included here is the increase in ionization rate and production of high-energy electrons due to the Maxwellization of electron energy distribution (Raizer, 1991, p. 229) when the electron density is high and electron-electron collisions become important. More detail is given in Appendix B. Maxwellization is usually not included in Monte Carlo simulations, such as those of Bakhov et al. (2000), due to numerical difficulties. We expect that, as the result of this process, the energy distribution will change from Druyvesteyn distribution, which has a small number of high-energy electrons, to Maxwellian, which has a higher number of high-energy electrons. In an electron-electron collision process, two electrons may combine their energies so that most of it goes into only one particle after collision. In such a manner, electrons can achieve energies higher than the potential difference that accelerated them, which is is impossible for electron-neutral collisions only in a static field. The electron-electron collision rate becomes comparable to electron-neutral collision rate at around $n_e \gtrsim 10^{21}$ m$^{-3}$, as estimated in Appendix B. This is of the same order of magnitude as $n_e$ at the streamer collision. The process of Maxwellization takes a finite time, several ps, which is comparable to the time of the existence of the high field during the streamer collision. Whether or not Maxwellization can play a role in production of high-energy electrons is a subject of future research.

Due to computing limitations, the streamers in the presented simulations were not allowed to grow to significant lengths. The short streamer length may affect some parameters in the streamer head, for example, it may limit its charge and radius. In particular, we did not yet model the branching of the streamers. One may argue that the pilots that create X-rays do not emerge from a small seed (as modeled here) but are rather initiated from existing channels of previous streamers, because these effects occur at a later stage of the streamer system development. The effects of the long streamer channels and branching will be also researched in the future.

### 4.3. Conclusions: Feasibility of the Fluid Model for Explanation of X-ray Emissions

We conclude that each streamer collision produces $\lesssim 10^3$ thermal runaway electrons of $\lesssim 10$-keV energy, but the probability of them to become relativistic runaways is low because the minimum energy required for electrons to be continuously accelerated by the calculated field is $\gtrsim 15$ keV. In the laboratory meter-scale and MV-voltage spark experiments usually around $10^4$ high-energy photons with typical energies $\sim 100$ keV are produced (Carlson et al., 2015; Kochkin et al., 2012, 2015). Let us estimate how many streamer collisions we need in a single spark to match this number, assuming that each collision produces a maximum possible estimated number, namely, $10^3$, of initial thermal runaway electrons, and they all become relativistic runaways. Let us also assume that they gain the maximum available electrostatic energy of $\sim 1$ MeV, which corresponds to the typical voltage used in the experiments. The probability of a 1 MeV to produce a photon >80 keV before slowing down may be estimated as the ratio of the photon production rate calculated using the bremsstrahlung cross section (Heitler, 1954, p. 245) and the slowdown rate due to effective friction force derived in Appendix C1 and is about 2%. Thus, the number of X-rays produced by each streamer collision is $\lesssim 20$. Thus, even with an optimistic assumption that all thermal runaways become relativistic runaways, the number of simultaneous streamer collisions required is at least $\sim 500$. However, the photo observations of streamer collisions only reveal a few ($\lesssim 10$) simultaneous streamer collisions (Kochkin et al., 2012, 2015).

There are also two additional objections to the presented theory. First, according to our test particle propagation modeling, the field conditions do not allow the thermal runaways to become relativistic. By the way, if we assume that the fraction of electrons that overcome the low electric field barrier is $\sim 10^{-3}$ (section 3.5), this gives the same order-of-magnitude answer of only 0.02 X-ray photons per streamer collision as PIC simulations of Köhn et al. (2017). Second, at least in the negative MV discharges, the photon energies may exceed the applied voltage (Kochkin et al., 2015). This is incompatible with the concept of electron acceleration by a static field. However, this is very important in relation to lightning, because the TGF may contain photons with very high energies (Briggs et al., 2010).

Thus, we arrive at the same conclusion as other authors (Babich & Bochkov, 2017; Ihaddadene & Celestin, 2015; Köhn et al., 2017) that the fluid model is insufficient to explain the observed X-ray emissions during streamer collisions. However, the experimental observations (Kochkin et al., 2012, 2015) suggest that X-ray photons are synchronized in time with the streamer collisions, so these phenomena must be linked (see section 1 for details on the experimental evidence).

The understanding of the exact scenario may involve considering additional mechanisms, as discussed in section 4.2. These mechanisms suggest inclusion of calculation of electron distribution over energies, that is, a kinetic or Monte Carlo/PIC approach (Chanrion & Neubert, 2008). One can combine fluid and particle approaches in hybrid models (e.g., Li et al., 2012).

Another conclusion is that low background ionization creates favorable conditions in the form of higher electric field and higher production of thermal runaway electrons. However, the rate of conversion of thermal runaways into relativistic runaways does not improve as the ratio of the maximum available energy for thermal runaway $\mathcal{E}_0$ and the minimum energy required for relativistic runaway, $\mathcal{E}_r$, stays about the same, $\mathcal{E}_0/\mathcal{E}_r \sim 0.5$–$0.7$.

## Appendix A: Photoionization

The photoionization process in air is production of photoelectrons from $O_2$ by ultraviolet photons within wavelength interval between 980 and 1,025 Å which are emitted by electron impact excited $N_2$ but still can propagate without being strongly absorbed by $N_2$ (Zheleznyak et al., 1982). It is described by a nonlocal source

$$s_p \equiv \left( \frac{\partial n_e(\mathbf{r})}{\partial t} \right)_{\text{photoionization}} = A \int s_i(\mathbf{r}') F(|\mathbf{r}' - \mathbf{r}|) \, d^3 \mathbf{r}'$$

where $s_i = \nu_i n_e$ is the impact ionization source. The kernel function $F$ is normalized to 1 and is given by expression

$$F_Z(r) = \frac{e^{-r/\Lambda_1} - e^{-r/\Lambda_2}}{4\pi r^3 \log(\Lambda_1/\Lambda_2)} \qquad (A1)$$

Here in the sea level air (pressure 760 mmHg, 20% content of $O_2$) $\Lambda_1 \approx 1.9 \times 10^{-3}$ m, $\Lambda_2 \approx 3.3 \times 10^{-5}$ m. This form of $F(r)$ was derived by Zheleznyak et al. (1982) by averaging the absorption by $O_2$ in the above mentioned wavelength interval, where it oscillates quickly (as a function of the photon wavelength) between the extreme values given by attenuation lengths $\Lambda_{1,2}$.

The normalizing coefficient $A$ is given by

$$A = A_0 \frac{p_q}{p + p_q} \qquad (A2)$$

where $p$ is the full air pressure (= 760 mmHg at sea level), $p_q$ is the parameter describing collisional quenching of excited molecules. We take $p_q = 60$ mmHg (Legler, 1963), while Zheleznyak et al. (1982) give the value $p_q = 30$ mmHg. The coefficient $A_0$ is given in the table of Zheleznyak et al. (1982) and varies in interval 0.05–0.12 in the electric field range of interest; we take a constant value of $A_0 = 0.1$.

### A1. Numerical Representation

For efficient computational purposes, the kernel function (A1) may be represented as a linear combination of functions (Luque et al., 2007)

$$G(r, \lambda) = \frac{e^{-r/\lambda}}{4\pi \lambda^2 r}$$

which are Green's functions of the Helmholtz equation

$$G - \lambda^2 \nabla^2 G = \delta(\mathbf{r})$$

We note that $G$ is also normalized to 1.

We follow Dubinova, (2016, ch. 9) and notice that (using the attenuation coefficient $\kappa = 1/\Lambda$)

$$F_Z(r) = \frac{1}{\log(\kappa_2/\kappa_1)} \int_{\kappa_1}^{\kappa_2} \frac{e^{-\kappa r}}{4\pi r^2} d\kappa = \frac{1}{\log(\Lambda_1/\Lambda_2)} \int_{\Lambda_2}^{\Lambda_1} \frac{1}{\Lambda} \frac{e^{-r/\Lambda}}{4\pi \Lambda r^2} d\Lambda$$

We note that this is also the way the formula was originally obtained by Zheleznyak et al. (1982) who averaged the attenuation over a range of wavelengths. The function under the integral can in turn be represented as

$$\frac{e^{-r/\Lambda}}{4\pi \Lambda r^2} = \frac{1}{\Lambda} \int_0^{\Lambda} G(r, \lambda) d\lambda$$

Combining these two expressions, we get

$$F_Z(r) = \frac{1}{\log(\Lambda_1/\Lambda_2)} \int_S \frac{G(r, \lambda)}{\Lambda^2} d\lambda d\Lambda$$

where $S$ is the area in $(\lambda, \Lambda)$ plane defined by $\Lambda_1 < \Lambda < \Lambda_2$ and $0 < \lambda < \Lambda$. Changing the order of integration to integrate over $\Lambda$ first, we finally get

$$F_Z(r) = \int_0^{\Lambda_1} W(\lambda) G(r, \lambda) d\lambda \qquad (A3)$$

where the weight function

$$W(\lambda) = \frac{1}{\log(\Lambda_1/\Lambda_2)} \left( \frac{1}{\max(\lambda, \Lambda_2)} - \frac{1}{\Lambda_1} \right)$$

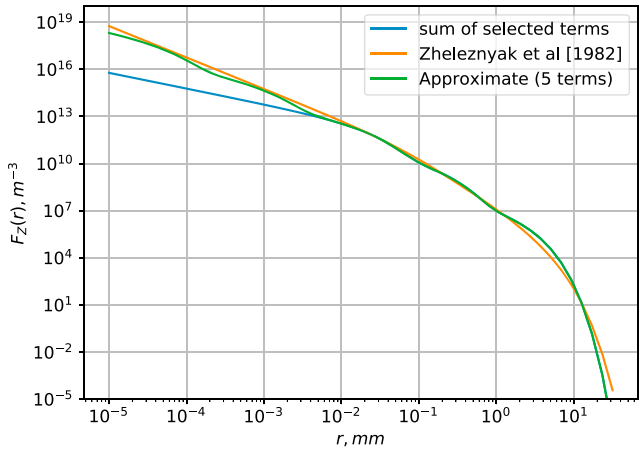

**Figure A1.** Approximation of Zheleznyak et al. (1982) photoionization kernel with a sum of Helmholtz equation solutions (A4) with $N = 5$ terms and the sum of terms $k = 3, 4, 5$ which was used in calculations.

Numerically, this may be represented as

$$F_Z(r) = \sum_{k=1}^{N} W_k G(r, \lambda_k) \quad (A4)$$

where $\lambda_k$ and $W_k$ are determined by $N$ intervals in $\lambda$ with boundaries $\lambda_{k+1/2}$, $k = 0 \ldots N$

$$\lambda_k = (\lambda_{k-1/2} + \lambda_{k+1/2})/2$$
$$W_k = \int_{\lambda_{k-1/2}}^{\lambda_{k+1/2}} W(\lambda)\, d\lambda$$

When choosing $\lambda_{k+1/2}$, we must make sure that all attenuation lengths are fairly represented. In particular, theory suggests that at least propagation of flat ionization fronts is determined by large $\lambda$ (Lehtinen et al., 2014), so the summation must extend as close to the upper limit $\Lambda_1$ as possible. The best choice seems to be logarithmically spaced $\lambda_k$, as demonstrated in Figure A1.

The values of $\lambda_k$ and weights $W_k$ are given in Table A1.

In practice, we found that only three terms are necessary for running simulations, namely, $k = 3, 4, 5$. The sum of selected terms is also plotted in Figure A1. This approximation accelerates computations and is justified by the following considerations:

1. The sum of the weights of the two neglected terms $W_1 + W_2 \approx 0.012 \ll 1$, so the total extra photoionization is small compared to the terms in use
2. The neglected terms contribute only at very short distances $\lesssim 1$ μm. This is of the order of our smallest grid step size, and numerically, the photoionization due to these terms is indistinguishable from small contributions to the impact ionization $\nu_i$ and electron diffusion $D$. This is seen from Helmholtz equation for the contribution $s_{pk}$ of term $k$ into total photoionization $s_p$:

$$s_{pk} - \lambda_k^2 \nabla^2 s_{pk} = W_k A s_i$$

If we neglect the spatial derivative term, then $s_{pk} \approx W_k A s_i$, substituting which back into this equation gives

$$s_{pk} = W_k A s_i + \nabla^2 \lambda_k^2 W_k A \nu_i n_e$$

By plugging this into equation (1), we see that this is equivalent to replace $\nu_i \to (1 + A \sum W_k)\nu_i$ and $D \to D + A\nu_i \sum W_k \lambda_k^2$ where the summation is over the neglected terms $k = 1, 2$. The resulting relative changes in $\nu_i$ and in $D$, respectively, do not exceed $10^{-4}$ and for $5 \times 10^{-4}$ for $E < 40$ MV/m.

### A2. Generalization to an Arbitrary Symmetric Kernel Function

This method may be generalized to an arbitrary symmetric kernel function $F(r)$ (see, e.g., Lagarkov & Rutkevich, 1994, pp. 85, 89) which satisfies certain integrability conditions (on which we will not focus here). We will now show that it can be represented as such a linear combination

$$F(r) = \int_0^\infty W(\kappa)G(r, \kappa)\, d\kappa, \qquad G(r, \kappa) = \frac{\kappa^2 e^{-\kappa r}}{4\pi r}$$

**Table A1**
*Coefficients for Approximating Zheleznyak et al. (1982) Photoionization Function (A4) With $N = 5$ Terms*

| k | $\lambda_k$, m | $W_k$ |
|---|---|---|
| 1 | $4.528 \times 10^{-8}$ | $6.197 \times 10^{-4}$ |
| 2 | $9.094 \times 10^{-7}$ | $1.183 \times 10^{-2}$ |
| 3 | $1.827 \times 10^{-5}$ | $2.376 \times 10^{-1}$ |
| 4 | $1.460 \times 10^{-4}$ | $4.662 \times 10^{-1}$ |
| 5 | $1.079 \times 10^{-3}$ | $2.838 \times 10^{-1}$ |

(we switched to using attenuation coefficient $\kappa = 1/\Lambda$). The derivation of $W(\kappa)$ is based on the idea that this looks almost like a Laplace transform. First, we notice that

$$\frac{e^{-\kappa r}}{r} = \int_{\kappa}^{\infty} e^{-\kappa' r} \, d\kappa'$$

Substitute this into $G$ and into the integral for $F(r)$ and switch the order of integration over $\kappa$ and $\kappa'$, being careful also about changing the limits

$$F(r) = \int_0^{\infty} e^{-\kappa' r} \, d\kappa' \int_0^{\kappa'} \frac{\kappa^2 W(\kappa)}{4\pi} \, d\kappa = \int_0^{\infty} e^{-\kappa' r} \tilde{F}(\kappa') \, d\kappa', \qquad \tilde{F}(\kappa) = \frac{1}{4\pi} \int_0^{\kappa} \kappa'^2 W(\kappa') \, d\kappa'$$

We see that (i) $\tilde{F}$ is the inverse Laplace transform of $F(r)$:

$$\tilde{F}(\kappa) = \frac{1}{2\pi i} \int_{\sigma - i\infty}^{\sigma + i\infty} e^{\kappa r} F(r) \, dr$$

and (ii) $W$ may be obtained from the expression for $\tilde{F}$

$$W(\kappa) = \frac{4\pi}{\kappa^2} \frac{d\tilde{F}}{d\kappa}$$

The final expression is thus

$$W(\kappa) = \frac{2}{i\kappa^2} \int_{\sigma - i\infty}^{\sigma + i\infty} F(r) e^{\kappa r} r \, dr$$

where $\sigma$ is chosen so that all the poles of $F(r)$ are to the left of the integration contour. If we prefer to work with attenuation lengths, as in equation (A3), then

$$W(\lambda) = -2i \int_{\sigma - i\infty}^{\sigma + i\infty} F(r) e^{r/\lambda} r \, dr$$

## Appendix B: Electron Maxwellization

The Maxwellization of electron energy distribution occurs when the energy loss rate in collisions with electrons exceeds that with neutrals, $\nu_{\mathcal{E}e} > \nu_{\mathcal{E}m}$. These may be defined as the respective momentum loss rates $\nu_{e,m}$ multiplied by the respective fractions of energy lost. The electrons lose a large fraction of their energy when colliding with another electron, so we can take $\nu_{\mathcal{E}e} = \nu_e$, while for collisions with molecules $\nu_{\mathcal{E}m} = \delta \nu_m$ where the fraction $\delta \sim 10^{-3} - 1$ depending on electron energy $\mathcal{E}_e = (3/2) T_e$ (Raizer, 1991, pp. 17–19), and $\nu_m$ is the electron momentum loss rate due to collisions with molecules.

The momentum loss rate due to collisions with electrons is $\nu_e = n_e \sigma_e v_t$, where $v_t = \sqrt{3 T_e / m_e}$ is the thermal electron velocity, and Coulomb cross-section $\sigma_e$ due to an electron-electron collisions is

$$\sigma_e = 4\pi b_0^2 \log \Lambda, \qquad \Lambda = \frac{\lambda_D}{b_0}, \qquad b_0 = \frac{e^2}{4\pi \varepsilon_0 \mathcal{E}_e}, \qquad \lambda_D = \sqrt{\frac{\varepsilon_0 T_e}{n_e e^2}}$$

where $\log \Lambda$ is the Coulomb logarithm, $b_0$ is the minimum target parameter, and $\lambda_D$ is the maximum target parameter, equal to the Debye length.

We can estimate $\mathcal{E}_e = (3/2) T_e$ as a function of electric field $E$ from the given values of mobility $\mu$ and diffusion coefficient $D$ (Morrow & Lowke, 1997) using Einstein relation $T_e = D/(e\mu)$. Moreover, $\delta$ may be estimated from relation $v_d/v_t \approx 0.8 \sqrt{\delta}$ (Raizer, 1991, equation (2.16)), where $v_d = \mu E$ is the drift velocity. The momentum loss rate due to collisions with molecules may be estimated from $\nu_m = e/(m_e \mu)$. Thus, $n_e$ at which $\nu_{\mathcal{E}e}$ exceeds $\nu_{\mathcal{E}m}$, that is, Maxwellization becomes important, is given by

$$n_e^{\text{maxw}} = \frac{\delta \nu_m}{\sigma_e v_t}$$

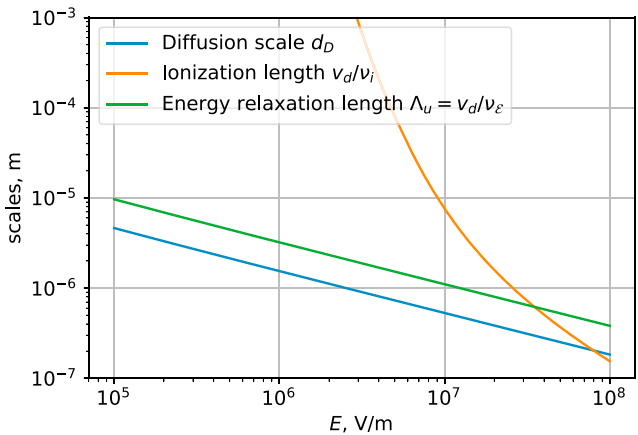

**Figure B1.** Various small scales important in air discharge at sea level.

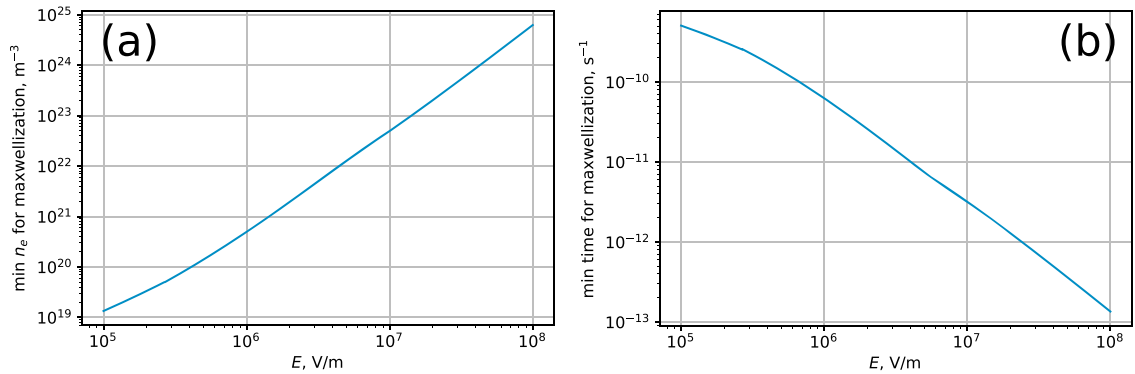

**Figure B2.** Requirements for Maxwellization of electron distribution: (a) Minimum electron density $n_e^{maxw}$. (b) Minimum time $1/\nu_e$.

and we estimate it to be $10^{21}$–$10^{22}$ m$^{-3}$ for $E = 1.5$–$4$ MV/m. The results of our calculations suggest that the density of of the order of $n_e = 10^{21}$ m$^{-3}$ may be achieved during the streamer collision process.

The Maxwellization may affect all of the reaction coefficient used in equations (1)–(3) due to change in the distribution, even though the average electron energy would not change. One may also question whether electron-electron collisions affect the coefficients of mobility $\mu_e$ and diffusion $D$. We can argue that there is no direct effect because the total momentum of two colliding electrons is conserved during a collision, so the average velocity (and therefore, $\mu_e$) is not affected (the same argument applies also for $D$). However, there is still an indirect effect through the change in collision frequency which is in turn due to change in the electron distribution.

In Figure B1, we plot the scales discussed in sections 4.1 and 4.2 and calculated as outlined above. The minimum electron density and time required for Maxwellization of electrons $1/\nu_e$ are plotted in Figure B2.

We point out that the process of electron-electron collisions is harder to model using Monte Carlo technique than electron-neutral collisions and therefore was not included in the model of Bakhov et al. (2000). If the electrons are totally Maxwellized, the fraction of electrons exceeding the thermal runaway threshold of $\mathcal{E}_t = 110$ eV is of the order of $e^{-\mathcal{E}_t/T_e} \approx 5 \times 10^{-5}$ for temperatures of the order $T_e = 10$ eV which are attained for the fields at streamer collision (Bakhov et al., 2000, Table 1).

## Appendix C: Representing Stochastic Motion of Electrons in a Deterministic Way

The electron motion in air is affected both by inelastic and elastic collisions. The inelastic collisions and incurred energy losses are conveniently described by the dynamic friction function $F_D(p)$, which is the time rate of loss of the absolute value of momentum, or stopping power (really, a force), given by Bethe formula (Bethe & Ashkin, 1953, p. 254). It is dependent on electron energy, or, equivalently, on the absolute value of momentum $p$. Electron motion is also affected by elastic collisions which change its direction. This process causes the loss of parallel momentum $\nu_m(p)p$, where $\nu_m$ is the elastic collision frequency, or momentum loss rate, and is also a function of $p$. To accurately describe the stochastic motion of electron which experiences collisions which change both the direction and value of its momentum, a full Monte Carlo calculation would be more appropriate. In the present work, we have a goal of a simple estimate of whether an electron can become a runaway or not, and thus, we will try to approximate the stochastic motion with a deterministic one by representing the effective friction force as some combination of $F_D$ and $\nu_m p$.

We follow a simple theory outlined by (Lehtinen et al., 1999, section 3.2), which assumes that elastic collisions are so frequent that the electron distribution function in momentum space assumes angular equilibrium much faster than equilibrium in absolute value of momentum. The distribution is given by equation (25) of Lehtinen et al. (1999), which, when rewritten with variables used in the present paper (we substitute the angular diffusion in terms of collision frequency, $D_{ang} = \nu_m/2$), becomes

$$f(\mathbf{p}, t) \propto f_p(p, t) \exp\left(\frac{2eE\cos\theta}{\nu_m p}\right)$$

where $\theta$ is the angle between the electron momentum **p** and the direction of electric force ($-e\mathbf{E}$), $f_p$ is the distribution in $p$, and the exponent represents the angular part of the distribution. From the kinetic equation, Lehtinen et al. (1999) then obtained that for an equilibrium at a given $p$, the equilibrium electric field $E_{eq}(p)$ is given by

$$eE_{eq}M(p, E_{eq}) = F_D(p) \tag{C1}$$

where

$$M(p, E) = \langle \cos\theta \rangle = \coth\xi - \frac{1}{\xi}, \quad \xi = \frac{2eE}{v_m p} \tag{C2}$$

is calculated from the angular part of distribution given above. The kinetic equation used by Lehtinen et al. (1999) suggests that the absolute value of momentum evolves approximately according to

$$\frac{dp}{dt} = eEM(p, E) - F_D(p)$$

which neglects spreading of distribution $f_p(p, t)$ due to stochasticity of collisions. The drift motion of electrons $\mathbf{v}_d = \dot{\mathbf{r}}$ is along $\hat{\mathbf{e}} = -\mathbf{E}/E$ and is related to the absolute value $v$ by

$$\mathbf{v}_d = \hat{\mathbf{e}} v M(p, E)$$

because of the angular spreading. For relativistic particles, the kinetic variables are related as

$$v = \frac{pc}{\sqrt{(mc)^2 + p^2}}$$

To summarize, the sought equations of motions are

$$\frac{dp}{dt} = eEM(p, E) - F_D(p) \tag{C3}$$

$$\frac{d\mathbf{r}}{dt} = -\frac{\mathbf{E}}{E} \frac{pcM(p, E)}{\sqrt{(mc)^2 + p^2}} \tag{C4}$$

where $M$ is given by (C2).

For comparison, when elastic collisions are completely neglected ($v_m = 0$), the equations of motion would be simply

$$\frac{d\mathbf{p}}{dt} = -e\mathbf{E} - \hat{\mathbf{p}} F_D(p) \tag{C5}$$

$$\frac{d\mathbf{r}}{dt} = \mathbf{v} \tag{C6}$$

### C1. The Effective Friction Force and Runaway Boundary

As mentioned above, there are several definitions of forces which contribute to slowing down of relativistic electrons:

1. The dynamic friction force $F_D(p)$ which describes losses of energy.
2. The parallel momentum loss rate $v_m(p)p$.
3. We can also define the effective friction which is equal to electric force $F_{eff} = eE_{eq}$ when equation (C3) gives equilibrium for $p$, that is, $dp/dt = 0$, and is given by $F_{eff}(p) = eE$, where $E$ is the solution of equation (C1) for given $p$.

If equation (C1) is solved for momentum $p$ for given $E$ instead, then the solution $p_s$ may be considered the approximate boundary of the runaway regime (i.e., electrons with $p > p_s$ become runaways while those with $p < p_s$ do not). As shown by Lehtinen et al. (1999), this approximation gives good agreement when the RREA rates are calculated, at least for not very high fields ($E \lesssim 1.5$ MV/m).

We plot various forces, namely, $F_D$, $v_m p$, and $F_{eff} = eE_{eq}$, in units of V/m at sea level air in Figure C1. While the minimum of $F_D$ is approximately 0.216 MV/m which occurs at $\mathcal{E}_D^{min} \approx 1.23$ MeV, the minimum of $F_{eff}$ is approximately 0.282 MV/m which occurs at $\mathcal{E}_{\text{eff}}^{\min} \approx 7$ MeV. Note that the latter must be considered the true runaway avalanche threshold field. It is close to the value of 0.284 MV/m calculated by Dwyer (2003) and 0.283 MV/m calculated by Babich et al. (2004). Moreover, Figure 4 in Babich et al. (2004) demonstrates that an initial energy of 10 MeV was required at this field for electrons to run away, which supports our conclusion that this energy is determined by $\mathcal{E}_{\text{eff}}^{\min}$ and not by $\mathcal{E}_D^{\min}$.

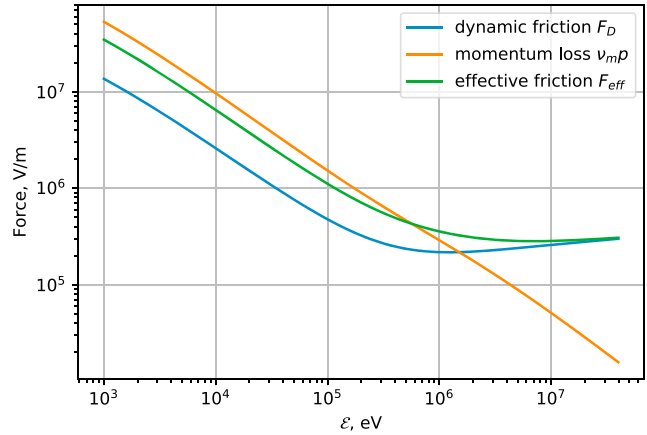

**Figure C1.** Various friction forces: the dynamic friction $F_D$ (loss of energy), the loss of parallel momentum $\nu_m(p)p$, and the effective friction $F_{\text{eff}}$ described in section C1. This effective friction force is approximately the one used in solving the equations of motion (C2)–(C4). If $F_{\text{eff}} = eE$, where field $E$ is given, then the corresponding energy is the approximate boundary of the runaway regime.

The increase in the field required for runaway compared to a theory without collisions is $F_{\text{eff}}/F_D \approx 2.5$ for electron energies 1 keV$< \mathcal{E} <$ 100 keV. Results of Bakhov et al. (2000) suggest that this ratio is somewhat smaller, ~1.35–1.55 at the peak of the dynamic friction function ($\mathcal{E} = 110$ eV, $F_D = 26$ MV/m) because they calculated substantial runaway transition at fields of 35–40 MV/m.

### C2. Checking Validity of Equations (C2)–(C4) in Asymptotic Cases

It is instructive to consider asymptotic cases as follows.

1. *Asymptotic case 1.* $\xi \gg 1$. Then $\nu_m p \ll eE$, and $M(p, E) \approx 1 - \frac{\nu_m p}{2eE}$ is close to 1. Physically, it means that electron moves almost parallel to $-\mathbf{E}$, deviating only by a small angle. This happens, for example, in the runaway regime. Introducing $\mathbf{p}_\parallel = \hat{\mathbf{e}} p M$, we get

$$\dot{\mathbf{p}}_\parallel = \hat{\mathbf{e}}(eEM^2 - F_D M) \approx -e\mathbf{E} - \nu_m \mathbf{p}_\parallel - \hat{\mathbf{p}} F_D$$

(we neglected the change in $F_D$, i.e., $F_D M \approx F_D$) which is the usual equation of motion with collisions and

$$\dot{\mathbf{r}} = \mathbf{v}_\parallel$$

where $\mathbf{v}_\parallel$ is calculated assuming particle momentum is $\mathbf{p}_\parallel$.

2. *Asymptotic case 2.* $\xi \ll 1$. Then $M(p, E) \approx \frac{2eE}{3\nu_m p} \ll 1$. Physically, it means that electron distribution is almost isotropic. This is valid for electrons that are not in the runaway regime. Such electrons are also usually nonrelativistic, which we will assume. The coordinate change is then given by the drift velocity

$$\dot{\mathbf{r}} = \hat{\mathbf{e}} \frac{2}{3} \frac{eE}{m\nu_m} = \mathbf{v}_d$$

and the momentum equation is

$$\dot{p} = \frac{2}{3} \frac{(eE)^2}{\nu_m p} - F_D(p)$$

which in equilibrium ($\dot{p} = 0$) gives

$$\sqrt{F_D \nu_m m v_t} = \sqrt{\frac{2}{3}} eE$$

where we substituted $p = mv_t$ because the equilibrium is achieved at some average ("thermal") velocity $v_t$. The dynamic friction may be represented in terms of energy loss rate $v_t F_D = \nu_\mathcal{E} \mathcal{E}_e = \delta \nu_m \mathcal{E}_e$ (for explanation of notations; see section B). From here

$$F_D = \frac{\delta \nu_m \mathcal{E}_e}{v_t} = \frac{\delta m \nu_m v_t}{2}$$

From the above equation we obtain the "thermal" velocity $v_t$

$$v_t = \sqrt{\frac{4}{3\delta}} \frac{eE}{m\nu_m} = \sqrt{\frac{3}{\delta}} v_d$$

which approximately coincides with the elementary theory of electron drift (Raizer, 1991, equation (2.16)). Note our use of quotes for "thermal" because the electron distribution is non-maxwellian.

**Acknowledgments**
This study was supported by the European Research Council under the European Union's Seventh Framework Programme (FP7/2007-2013)/ERC grant agreement n. 320839 and the Research Council of Norway under contracts 208028/F50, 216872/F50 and 223252/F50 (CoE). The full source of the code described here is available at https://git.app.uib.no/Nikolai.Lehtinen/streamer_2d_cyl.

**Journal of Geophysical Research: Atmospheres**  10.1029/2018JD028646